\documentclass[10pt,letterpaper]{article}
\usepackage[top=0.85in,left=2.75in,footskip=0.75in,marginparwidth=2in]{geometry}

\usepackage[utf8]{inputenc}
\usepackage{ragged2e}
\usepackage{cite}

\usepackage{nameref,hyperref}

\usepackage{microtype}
\DisableLigatures[f]{encoding = *, family = * }

\raggedright
\usepackage{changepage}

\usepackage[aboveskip=1pt,labelfont=bf,labelsep=period,singlelinecheck=off]{caption}

\makeatletter
\renewcommand{\@biblabel}[1]{\quad#1.}
\makeatother

\usepackage{lastpage,fancyhdr,graphicx}
\usepackage{epstopdf}
\fancyheadoffset[L]{2.25in}
\fancyfootoffset[L]{2.25in}

\usepackage{color}

\definecolor{Gray}{gray}{.25}

\usepackage{graphicx}

\usepackage{sidecap}

\usepackage{wrapfig}
\usepackage[pscoord]{eso-pic}
\usepackage[fulladjust]{marginnote}
\reversemarginpar
\usepackage{epstopdf}
\usepackage{multirow}
\usepackage{relsize}
\usepackage{xpatch}
\usepackage{nccmath}
\usepackage{amsmath}

\DeclareMathOperator*{\argmin}{argmin}

\begin{document}
\vspace*{0.35in}

\begin{flushleft}
{\Large
\textbf\newline{ Image Reconstruction from Undersampled Confocal Microscopy Data using  Multiresolution Based  Maximum Entropy Regularization\footnote{Submitted to IOP JINST}}
}
\newline
\\
Bibin Francis\textsuperscript{1},
Manoj Mathew\textsuperscript{2},
Muthuvel Arigovindan\textsuperscript{1},
\\
\bigskip
\bf{1} Imaging Systems Lab, Electrical Engineering, Indian Institute of Science (IISc), Bangalore, 560012, India
\\
\bf{2}Central Imaging \& Flow Cytometry Facility, National Center for Biological Sciences (NCBS), Bangalore 560065, India
\\
\bigskip
{Corresponding author: mvel@iisc.ac.in}

\end{flushleft}

\section*{Abstract}
\justify
We consider the problem of reconstructing 2D images from randomly under-sampled
confocal microscopy samples. The well known  and widely celebrated total variation 
regularization, which is the $\ell_1$ norm  of derivatives, turns out to be unsuitable for 
this problem; it is unable to handle both noise and under-sampling  together.
 This issue is linked  with the notion of phase transition phenomenon 
observed in {\em compressive sensing} research, which is essentially  the break-down 
of total variation methods, when sampling density gets lower than certain threshold.  
The severity of this breakdown is  determined by the so-called {\em mutual incoherence} between 
the derivative operators and measurement  operator. In our problem, the mutual incoherence
 is low,  and hence the total variation regularization gives serious  artifacts in the presence of 
 noise even when the sampling density is not very low.  There has been very few attempts in
 developing   regularization methods that perform better than total variation  regularization for 
 this problem.  We develop a    multi-resolution based regularization method that is adaptive to
 image structure.    In our approach,   the desired  reconstruction is  formulated  as a series of 
 coarse-to-fine  multi-resolution reconstructions;  for reconstruction at each level,  the regularization
  is constructed to be adaptive to the image structure, where the information for adaption is 
  obtained from the reconstruction obtained at coarser resolution level.  This adaptation  is 
  achieved by using maximum entropy principle, where the required adaptive regularization is 
  determined as the maximizer of entropy subject to the information extracted from the coarse  
  reconstruction as constraints.  We also utilize the directionally adaptive second order derivatives 
  for constructing the regularization  with directions guided by the given coarse  reconstruction,  
  which leads to an improved suppression of artifacts.  We demonstrate the superiority of the
   proposed regularization method over existing ones using several reconstruction examples.


\section{Introduction}
Confocal microscopy is a wide-spread tool among cellular biologists for studying functionality and physiology of living cells,  and it has a theoretical resolution that is better than that of the  wide-field microscopy.  Its effective resolution becomes comparable to widefield microscopy, or even inferior to that of wide-field microscopy in the case of the long term live cell imaging because of the noise.  However, it is still preferred  over widefield microscopy by many researchers for the   possible reason that  the raw images can be interpreted directly without any deconvolution. This means that it can be applied in 2D+t  mode for observing fast cellular phenomena~\cite{gonzalez2002real, sokolov2003real}.

In this viewpoint, the frame rate in  confocal microscopy  is limited because of
the point-wise scanning.   Most of the confocal microscopes take  $0.1 - 1$ seconds to generate a single 2D image~\cite{wnek2008confocal}. Moreover, the point wise scanning coupled with the pinhole screen in confocal microscope significantly reduces the number photons impinging on the detector. As a result, the signal-to-noise ratio (SNR) of the reconstructed image is low. To compensate for this, the excitation intensity should be raised which will lead to photo-bleaching and photo-toxicity. Photo-bleaching is the process by which the dye combines with the atmospheric oxygen and becomes non-fluorescent. In practical cases, photo-bleaching is reduced by reducing the level of oxygen in the system. Photo-toxicity is the process by which dye reacts with the living cell and in turn destroying it. The only way to control the  photo-toxicity is to compromise on the resolution by either by reducing the excitation intensity,  or by increasing the pinhole size;  while the former approach leads to loss of resolution by means of increased noise,  the latter leads to loss of resolution in the form of increased blurring.

Considering the problem of longer acquisition time, we find that there has been a significant progresses such as the use of acoustic-optic deflector (AOD)~\cite{tsien1995video,iyer2003compensation}, and the development of Nipkow disc~\cite{xiao1988real}; however the point scanning always limits the acquisition speed. Our goal is to develop a computational tool that will allow trading-off resolution  against  acquisition  time. To achieve this,  we develop a method to reconstruct a full 2D images from randomly sub-sampled confocal measurements.  The domain of scientific computing that deals with this problem is called the scattered data approximation (SDA). Scattered data approximation methods are broadly classified as mesh-based methods~\cite{lai1996scattered}, Krigging-based methods~\cite{stein2012interpolation}, distance-weighted methods~\cite{shepard1968two},  polynomial approximation methods~\cite{fenn2006robust}, and roughness minimizing methods~\cite{duchon1977splines,micchelli1984interpolation}. Roughness minimizing methods have better robustness to noise and fluctuations in sampling density. So, we restrict our survey to the roughness minimizing methods.

Given the list of scattered sample locations $\{\mathbf{x}_i\}_{i=1}^{N_s}$, and corresponding sample values $\{{f}_i\}_{i=1}^{N_s}$, roughness minimizing method addresses the reconstruction problem as the following minimization problem, 
\begin{equation}
\label{eq:rmrec}
u^* = \argmin_{u}\sum_{i=1}^{N_s}\left(u(\mathbf{x}_i)-f_i\right)^2  + \lambda R(u),
\end{equation}
where $R(u)$ is the  roughness functional.  As the sample locations $\{\mathbf{x}_i\}_{i=1}^{N_s}$ can be arbitrary, the mathematically correct way to handle the above minimization problem is to search for minimum with a suitable space of continuous functions (not a discrete image),  as done by the thin plate spline methods. The landmark papers~\cite{duchon1977splines, micchelli1984interpolation} analytically solve the above minimization with some reasonable assumptions on the search space with the following form is regularization functional:
\begin{equation}
\label{eq:roughness1}
R(u) = 
	\int\int\left[\left(\frac{\partial^2 u(\bf{x})}{\partial x^2}\right)^2+
	2\left(\frac{\partial^2 u(\bf{x})}{\partial xy}\right)^2+
	\left(\frac{\partial^2 u(\bf{x})}{\partial y^2}\right)^2\right]dxdy.
\end{equation}
Solving the above mathematical minimization problem leads to the following form of the solution $u^*$ as given below:
\begin{equation}
\label{eq:rbfsoln}
u^*({\bf{x}}) = \sum_{i=1}^{N_s}w_i\phi(\|{\bf{x}}-{\bf x}_i\|_2) + p({\bf x}),
\end{equation}
where $\phi(\bf{x})$ is the thin-plate spline (TPS), and $p({\bf x})$ is a degree-one  2D polynomial. Although this method is mathematically elegant, it suffers from the following problem: (i) the weights  have to be computed by solving a ill-conditioned, dense, large system of equations  which breaks down numerically when the number of samples, $N_s$ becomes higher few thousands;  (ii) after computing the weights, the required regular image is not directly available, and it has to be computed from the equation (\ref{eq:rbfsoln}), which is an expensive operation. The first problem is somewhat addressed by Radial basis function method~\cite{buhmann2003radial,carr1997surface}, and is fully eliminated by the partition of unity method~\cite{cavoretto2015partition}  by slightly compromising on the reconstruction quality. The subspace variational method~\cite{arigovindan2005variational,morozov2011reconstruction} eliminates both problems present in the TPS method; however,  it requires solving a linear system of equations of size that is equal to the number of pixels in the final reconstructed image, which may require a large amount of memory,  and may require specially built sparse solvers.

To eliminate the need for storing large matrices, we developed a discrete formulation in~\cite{francis2018scattered}. First, the sets $\{\mathbf{x}_i\}_{i=1}^{N_s}$ and  $\{{f}_i\}_{i=1}^{N_s}$ are transformed into
a pair of regular grid images as follows:
\begin{align}
\label{eq:chdef1}
c(\mathbf{y})   & = \sum_{i\in {\cal N}_{\bf y}} w_i(\mathbf{y}), \;\;\;\; w_i(\mathbf{y}) = \frac{tanh(\|\mathbf{y}-\mathbf{x}_i\|)}{\|\mathbf{y}-\mathbf{x}_i\|}  \\
\label{eq:chdef2}
h(\mathbf{y}) &  =  h^{\prime}(\mathbf{y})/c(\mathbf{y}), \;\; h^{\prime}(\mathbf{y})   = \sum_{i\in {\cal N}_{\bf y}} w_i(\mathbf{y})f_i, 
\end{align}
where ${\cal N}_{\bf y}$  is set of indices such that the corresponding sample locations are within the unit area neighborhood around $\mathbf{y}$.

With this,  the possible  reconstruction methods that we intend to analyze can be collectively 
written as follows: 
\begin{align}
\label{eq:regrecsda}
u^* = \argmin_{u} \sum_{\mathbf{y}}c(\mathbf{y})\left(h(\mathbf{y}) - u(\mathbf{y})\right)^2 
 + \lambda\underbrace{ \sum_{\mathbf{y}}\left(\sum_{i=1}^3 \left(\left(d_i\ast u\right)
		(\mathbf{y})\right)^2\right)^{p/2}}_{R_p(u)},
\end{align}
where $\{d_i,i=1,2,3\}$'s are discrete filters implementing  derivative operators
$\frac{\partial}{\partial x\partial x}, \frac{\partial^2}{\partial y\partial y}, \\ 
\sqrt{2}\frac{\partial}{\partial x\partial y}$. The actual cost solved in~\cite{francis2018scattered} is with $p=2$, where  we demonstrated that reconstruction obtained by minimizing the above cost with $p=2$ reproduces the mathematically exact solution (equations \eqref{eq:rmrec}, and \eqref{eq:roughness1})  with good accuracy. Further, minimizing the above cost does not require storing large matrices and only requires iterations involving filtering and array multiplications only. The method presented in~\cite{unser1995multigrid}  is special case of our  work presented in~\cite{francis2018scattered}, where the sample locations $\mathcal{X}$ are restricted to in a subset of regular grid points.  

One drawback in reconstructing the required image via solving~\eqref{eq:regrecsda} with $p=2$ is that it  penalizes the square of the roughness at each pixel  locations,  which causes over-smoothing  of sharp region of the image.  This is clearly undesirable for image reconstruction in general, and fluorescence image recovery in particular.  Fluorescent images are made of piecewise smooth regions separated by  sharp intensity changes. In fluorescent imaging, the part of the specimen to be imaged is injected with a dye. Upon the excitation of the specimen with a high intensity laser beam, only the dyed region emit light and hence they appear bright compared to other regions in the image. As a result of this, the fluorescent images have sharp edges,  and hence quadratic regularization is not suitable.

In general image recovery problems  such as   deblurring of  full  fluorescence images~\cite{scherzer1998denoising}, reconstruction  from Fourier samples  measured using magnetic resonance~\cite{lustig2007sparse}, tomographic reconstruction~\cite{zheng2017sparse}, it has been proven that solving~\eqref{eq:regrecsda} with $p=1$---in which case the regularization is called the $\ell_1$ regularization---gives reconstruction with higher sharpness and resolution.  The main reason for this improvement is that minimizing the absolute value of derivatives by setting $p=1$ is that it allows a few large values of derivative magnitudes, while the quadratic roughness functional  forbids any large derivative magnitude. Hence the regularization with $p=1$ allows sharp edges in the reconstruction. The difference in performance between these forms of regularization functionals corresponding to $p=1$ and $p=2$  is even more pronounced when the noise is high. For  $p < 1$, it has been demonstrated that we get further improved performance over the quadratic regularization functional as well as the $\ell_1$ regularization functional in inverse problems such as deblurring~\cite{nikolova2010fast}, and tomographic reconstruction~\cite{Miao_2016}. Interestingly, there are no reported method that use other than $p=2$ for the current problem, i.e., the problem of  reconstructing  scattered spatial point measurements. In reconstruction trials involving minimization of \eqref{eq:regrecsda} with $p=1$, we observed spike-like artifacts that appear to be a kind of  remnants of sample density distribution. It was also reported that the reconstructions obtained  with $p \in (0,1]$ were actually worse than the ones obtained with $p=2$  in terms  structural similarity measures~\cite{francis2018multiresolution}.

A related problem is observed in inverse problems such as the reconstruction  from Fourier samples,tomographic reconstruction, where the image reconstruction  using the regularization with $p=1$ breaks down leading to reconstruction quality that is  significantly inferior to the quality yielded by quadratic case, i.e., the case with $p=2$. This effect is referred in the literature as phase transitions~\cite{donoho2011noise}. The under-sampling  factor at which the reconstruction breaks down is determined by the so-called mutual incoherence \cite{candes2007sparsity} between the operator involved in measurement and derivative operator used to construct the regularization. The actual definition of mutual incoherence is beyond the scope of this paper. However, it is sufficient to note that, for the problem addressed in the paper,  i.e., the reconstruction from spatial point measurements, mutual incoherence is the lowest. Hence the phase-transition threshold  on the sample density is significantly higher,  and hence the occurrence of spikes mentioned above is justified.

A novel method, named as anisotropic interpolation method,  for  reconstructing  images from sparse, spatial point measurement has been developed in~\cite{bourquard2013anisotropic}. Here the reconstruction is achieved by means of a series of quadratic regularized reconstructions, where each quadratic regularization is constructed using first order derivatives such that it is directionally adaptive by making use of the information from the previous reconstruction in the series.  However,  this method is based on first order derivatives and hence is not suitable for fluorescence image, as the fluorescence images are composed of sharp structures. In~\cite{francis2018multiresolution}, we developed a new multi-resolution based regularization method using probabilistic formulation, where the required reconstruction is obtained by solving a series of reconstructions at different resolution. For the reconstruction at each resolution level,  the regularization is constructed using a probabilistic view point such that it is adaptive to the local image structure;  the local image information is obtained from the previous reconstruction in the series. We named this method as the MSDA method,  and we demonstrated that MSDA  outperforms both $l_1$ method (solving \eqref{eq:regrecsda} with $p=1$) and anisotropic method of \cite{bourquard2013anisotropic}.

In this paper, we develop a further significantly improved regularization method for image reconstruction from non-uniformly under-sampled spatial point measurements. The current formulation is better than the previous one~\cite{francis2018multiresolution} in the following ways: (i) it is designed to be adaptive to local image structure based on maximum entropy principle;  (ii) it utilizes image Hessian to construct  directionally adaptive regularization designed to further reduce  the  distortions.  In the next section,  we provide  a more detailed  overview of existing methods  by introducing the formulation of  maximum a posteriori estimation; this will serve for introducing the notions required for the development of the proposed method. In section 3,   we develop   our Hessian based maximum  entropic regularization that is constructed based on the given  prior estimate of the required solution.  We also a introduce the overall cost function that will lead to the  proposed reconstruction method. In Section 4,  we describe the multiresolution approach for  eliminating the need for  prior estimate.   In Section 5, we provide experimental validation of the proposed method,  where we demonstrate that the proposed method outperforms existing methods with significant reduction in the amount of artifacts. 

\section{Regularized Reconstruction as MAP Estimation}
\label{Sect:RegMAPEst}

We will first describe the standard MAP approach that corresponds to the reconstruction problem given in the equation \eqref{eq:regrecsda}, and then will describe the modifications that will lead to the MSDA method. Let  $\mathcal{X}$ denote the set of sample locations, i.e.,  let  $\mathcal{X}= \left\{\mathbf{x}_1,\mathbf{x}_2\ldots\mathbf{x}_{N_s}\right\}$, and let $\mathbf{f}= \left[f_1,f_2,\ldots,f_{N_s}\right]^T$ denote the vector containing samples measured from the locations contained in the set  $\mathcal{X}$.   Let $h$ and $c$ be the image pair representing the measurement transformed using the rules given in the equations~\eqref{eq:chdef1} and~\eqref{eq:chdef2}.  Note that $h$ plays the role of $\mathbf{f}$ and $c$ plays the role of  $\mathcal{X}$.  For a candidate image, $u$, its probability of being the source of measurement, $h$, is known as the posterior probability, which can be expressed using Bayes rule as
\begin{equation}
\label{Eq:MAPFor}
p(u\vert h,c) = \frac{p(h\vert u,c)
	p(u,k)}{p(h,c)},
\end{equation}
where $p(h\vert u,c)$ is the probability of obtaining the measurement image $h$ given the candidate $u$ and the sample density map $c$,  $p(h,c)$ is the probability  of obtaining the measurement image $h$ given sampling density map $c$, and $p(u,k)$ is prior probability with variance parameter $k$.  In MAP approach, one determines the required image as the minimizer of the negative logarithm of $p(u\vert h,c)$.  Denoting $J(u,h,c) =-\log p(u\vert h,c)$, we write 
\begin{equation}
\label{Eq:MAPcost1}
J(u,h,c) =
\left[
{-\log p(h \vert u,c)}
-\log p(u,k)\right],
\end{equation} 
where we have ignored the term $p(h, c)$ since it is independent of $u$. The form of prior probability that will lead to the most known forms of regularization functionals can be written as
\begin{equation}
\label{eq:probform}
p(u,k)  = \prod_{\bf y}exp(-\frac{1}{k} (E_{[u]}({\bf y}))^{p/2}),
\end{equation}
where $E_{[u]}({\bf y})$ is some roughness image obtained from the candidate image $u({\bf y})$, typically in the form of sum of squares of derivative chosen order. Using the above form in the equation (\ref{Eq:MAPcost1}), gives
\begin{equation}
\label{Eq:MAPcost2}
J(u,h,c) =   D(u,h,c) + 
\frac{1}{k} \sum_{\bf y}(E_{[u]}({\bf y}))^{p/2} ,
\end{equation} 
where $D(u,h,c) = -\log p(h \vert u,c)$, which is usually called
the data fitting term.  Note that $p(h \vert u,c)$ is only determined by the random process  that generates the noise in the original measurement vector $\mathbf{f}$. Noise across the different components are clearly independent;  however, the noise across different pixels of $h$ are independent only if, for each $\mathbf{y}$,  the pixel value of $h$ comes from only one of the component of  $\mathbf{f}$.  This will be  true if the measurement sample density  is smaller than the reconstruction grid density, which is the case in the problem we are addressing in this paper. With this assumption,  and with assumption that the  noise is Gaussian, $D(u,h,c)$ is written as 
\begin{equation}
\label{eq:duhc}
D(u,h,c) = \frac{1}{\sigma^2}\sum_{\mathbf{y}}c(\mathbf{y})\left(h(\mathbf{y}) - u(\mathbf{y})\right)^2,
\end{equation}
where $\sigma^2$ is the noise variance. Although the noise in fluorescence imaging is not strictly Gaussian,  the above form is used for data fitting,  because of its low computational complexity.  It is customary to allocate higher computational complexity for the prior, $p(u,k)$ because the pay-off is more,  and hence the above form for $D(u,h,c)$ is frequently used. Substituting this form in \eqref{Eq:MAPcost2} gives
\begin{equation}
\label{Eq:MAPcost3}
J(u,h,c) =   
	\frac{1}{\sigma^2} \sum_{\mathbf{y}}c(\mathbf{y})\left(h(\mathbf{y}) - u(\mathbf{y})\right)^2
	+ \frac{1}{k} \sum_{\bf y}(E_{[u]}({\bf y}))^{p/2}.
\end{equation} 
The above cost  becomes identical upto a scale factor,  to that of  minimization given  in~\eqref{eq:regrecsda} with the substitution $E_{[u]}({\bf y}) = \sum_{i=1}^3 \left(\left(d_i\ast u\right)(\mathbf{y})\right)^2$ and $\lambda = \frac{\sigma^2}{k}$.  Since both $\sigma^2$  and $k$ are usually unknown,  $\lambda$ is typically considered as an user parameter.

Now,  we can see how the MAP formulation helps to explain the reason for getting spiky artifacts that appear in the reconstruction obtained by minimizing \eqref{Eq:MAPcost3} with $p=1$. From \eqref{eq:probform}, it is clear that the   reconstruction obtained as the minimizer of~\eqref{Eq:MAPcost3}  is based on the assumption  that the roughness value at each pixel location is independent of neighboring pixel location,  which is obviously wrong. Since both the operators involved in the cost---the derivative operator in the regularization and the Dirac delta in the data fitting part---are localized, ignoring dependency  of the roughness at a pixel location to that of its neighbors,  causes the spikes to appear in the reconstruction under the sparsifying effect of setting $p=1$. This does not occurs when $p=2$, because in this case,  minimizing $R_p$ does not force the derivatives to be sparse.  This idea might have some interesting connections with the idea of phase transitions~\cite{donoho2011noise}  analyzed in compressive sensing theory.

To mitigate the effect this problem,  we proposed the following modification to the prior probability $p(u,k)$. Suppose $v$ denotes the lower resolution estimate of the original image that generated the measurement.  We use this  as a guide for determining correction to be applied for compensating the error incurred by the pixel-wise independent assumption of the image roughness.  Then, we replace 
$p(u,k)$ with the following form of probability:
\begin{equation}
\label{Eq:msdaprior}
p(u, v, k) = \prod_{\mathbf{y}}\exp\left(
	- \frac{1}{k} \left( E_{[v]}(\mathbf{y})\right)^{-q}
	\left(E_{[u]}(\mathbf{y})\right)^r
	\right)
\end{equation} 
Clearly here,   $E_{[v]}(\mathbf{y})$  gives the  weights   for compensating for  the error incurred by the pixel-wise independent assumption. As evident, the weight is essentially the reciprocal of the total roughness in the lower resolution image $v$.  This is justified because,  if the average roughness around a pixel $\mathbf{y}$  in low resolution estimate, $v$, is high,  then the probability that the roughness being high in the required image $u^*$  at location $\mathbf{y}$ will also be higher. Now we rewrite~\eqref{Eq:MAPcost1}  using \eqref{Eq:msdaprior}  with $p(u,k)$  replaced by $p(u, v, k)$,  and with $-\log p(h \vert u,c)$   replaced    $D(u,h,c)$ to get the following
expression:
\begin{align}
\label{Eq:MAPcost4}
J_{msda}(u, v, h,c) &=  \frac{1}{\sigma^2}
	\sum_{\mathbf{y}}c(\mathbf{y})\left(h(\mathbf{y}) - u(\mathbf{y})\right)^2
\frac{1}{k} \underbrace{\sum_{\bf y}(E_{[u]}({\bf y}))^{-q}(E_{[u]}({\bf y}))^{r}}_{R_{msda}(u,v)}
\end{align}
 
 Next, note that $v$ is itself an unknown because it depends on the original
 image. We used a multi-resolution method to resolve this dependency problem.   To describe the multi-resolution approach,  let ${\cal L}_{[j]}$ denote a $2^j$-fold 2D interpolation operator such that  ${\cal L}_{[j]}u$   is an image whose size is $2^j$ larger than that $u$.  Let $J^{(j)}_{msda}(u, v, h, c)=J({\cal L}_{[j]}u, v, h, c)$ be the cost given in the equation~\eqref{Eq:MAPcost4} applied on $u$ after interpolating by the factor $2^j$. This means that, if $N\times N$ is the size of $h$  and $c$,  then $u$  in $J^{(j)}_{msda}(u, v, h, c)$  denotes an $\frac{N}{2^j}\times \frac{N}{2^j}$ variable.  Let $J^{(j)}(u, h, c)$ denote the cost derived from the cost of equation \eqref{Eq:MAPcost3} by the same way. Let $s$ denote the number of levels for implementing the multi-resolution. In the multi-resolution approach, we obtain the initializing reconstruction by the following minimization:
 \begin{equation}
 \label{eq:1mrr1}
 u^{(s)} = \underset{u}{\operatorname{argmin}} \;\; J^{(s)}(u, h, c)
 \end{equation}
 Using $u^{(s)}$ as initialization,  multi-resolution based reconstruction involves a series
 of minimizations for $j= s-1,s-2,\ldots,0$
 as given below:
 \begin{eqnarray}
 \label{eq:1mrr2}
 v  = & {\cal L}_{[j+1]}u^{(j+1)} \\
 \label{eq:1mrr3}
 u^{(j)}  = & \underset{u}{\operatorname{argmin}} \;\; J^{(j)}_{msda}(u, v, h, c)
 \end{eqnarray}
 Note that for $j=0$, the final minimization,   $u^{(0)}  = \underset{u}{\operatorname{argmin}} \;\;  J^{(0)}_{msda}(u,v,h, c)$ gives the required  final reconstruction.  Note that,  in the multi-resolution loop,   the result of previous reconstruction  play the role of $v$,  which is the guide to determine the  compensating weight for the current reconstruction. We demonstrated experimentally in~\cite{francis2018multiresolution}   that the MSDA method outperforms the $\ell_1$ method and anisotropic interpolation method of~\cite{bourquard2013anisotropic}. Further, the results of MSDA was observed to have less artifacts.

Although MSDA performed better than  $\ell_1$ method and anisotropic method,  we  still found some artifacts in the reconstructions, albeit  the amount of artifacts is significantly less. A possible discrepancy that could cause this artifact is that there is some level of arbitrariness in the compensation for the error incurred by pixel-wise independence assumption on the roughness. Specifically, in the form of  probability given in the \eqref{Eq:msdaprior}, $E_{[v]}(\mathbf{y})$  works as the averaged value   for $E_{[u]}(\mathbf{y})$ to be used for the compensating weight. This idea is actually put to work in the multiresolution loop specified by the equations \eqref{eq:1mrr2}  and \eqref{eq:1mrr3}. For reconstruction at scale $i$, we minimize $J^{(i)}_{msda}(u, v, h, c)$, containing  the regularization $R_{msda}({\cal L}_{[i]}u,v)$ which is defined on ${\cal L}_{[i]}u$ with compensating weights extracted from  $v={\cal L}_{[i+1]}u^{*(i+1)}$. Clearly ${\cal L}_{[i+1]}u^{*(i+1)}$ is some sort of averaged version of the  minimizer of $J^{(i)}_{msa}(u, v, h, c)$,  which is  ${\cal L}_{[i]}u^{*(i)}$.  However, we do not have a direct relation between them,  and hence the following question is left unanswered: what is the optimal amount of averaging required for defining the compensating weight? The averaging is implicitly  determined by the interpolation filter used in the definition of ${\cal L}_{[i]}$'s. This arbitrariness is probably the reason for the artifacts observed in the reconstruction of MSDA. 

\section{The proposed method: Maximum entropic regularized reconstruction}
\subsection{Hessian and directional derivatives}
Our construction of  regularization was inspired from  Hessian-Schatten norm introduced by Lefkimmiatis et al.~\cite{lefkimmiatis2013hessian}. The Hessian operator for continous function is given by
\begin{equation}
\nabla^2 = \left[
\begin{array}{cc}
\frac{\partial^2}{\partial x^2} & \frac{\partial^2}{\partial x\partial y} \\
\frac{\partial^2}{\partial x\partial y} & \frac{\partial^2}{\partial y^2}
\end{array}
\right].
\end{equation}
The Hessian of a 2D function is useful for computing directional second derivatives. Second derivative of a function $g(\mathbf{y})$ at point $\mathbf{y}$ along a direction $\mathbf{d}$ is defined as  $\frac{\partial^2}{\partial \alpha^2}g(\mathbf{y}+\alpha \mathbf{d})\vert_{\alpha=0}$, which, for brevity we denote by 
$\mathlarger{\mathlarger{\partial}}_{\mathbf{d} \mathbf{d}}$.   It can be shown that $\mathlarger{\mathlarger{\partial}}_{\mathbf{d} \mathbf{d}}g(\mathbf{y}) = \mathbf{d}^T\nabla^2g(\mathbf{y})\mathbf{d}$. For any two linearly independent direction vectors $\mathbf{d}_1$ and $\mathbf{d}_2$,  we can define  mixed directional derivative as $\frac{\partial^2}{\partial \alpha_1\partial \alpha_2}g(\mathbf{y}+\alpha_1 \mathbf{d}_1 + \alpha_2\mathbf{d}_2)\vert_{\alpha_1=0,\alpha_2=0}$, which we denote by $\mathlarger{\mathlarger{\partial}}_{\mathbf{d}_1 \mathbf{d}_2}$.  It can be shown that this becomes equal to$\mathlarger {\mathlarger {\partial} }_{\mathbf{d}_1 \mathbf{d}_2}g(\mathbf{y})=\mathbf{d}^T_1\nabla^2 g(\mathbf{y})\mathbf{d}_2$. The  discrete equivalent of Hessian that can be applied on discrete images can be expressed as
\begin{equation}
\mathbf{H}(\mathbf{y})=\left[ \begin{array}{cc} d_{xx}(\mathbf{y}) & d_{xy}(\mathbf{y}) \\
d_{xy}(\mathbf{y}) & d_{yy}(\mathbf{y})
\end{array}\right],
\end{equation}
where  $d_{xx}$, $d_{yy}$ and  $d_{xy}$ are discrete filters implementing the corresponding derivatives. We use the same notations for the discrete directional derivatives, and write  
$\mathlarger{\mathlarger{\partial}}_{\mathbf{d} \mathbf{d}}g(\mathbf{y}) =\mathbf{d}^T \left[(\mathbf{H}\ast g)(\mathbf{y})\right]\mathbf{d}$,  and
$ \mathlarger {\mathlarger{\partial}}_{\mathbf{d}_1
		\mathbf{d}_2} g(\mathbf{y}) = \mathbf{d}^T_1 \left[(\mathbf{H}\ast g)(\mathbf{y})\right]\mathbf{d}_2$. Here the notation $(\mathbf{H}\ast g)(\mathbf{y})$ represents convolving the matrix filter $\mathbf{H}$ with image  $g$ to get  a  $2\times 2$  matrix of images, and then accessing the matrix corresponding to the pixel location $\mathbf{y}$. The Hessian-Schatten norm is actually a family of norms and we present only a restricted, but most useful form below:
\begin{equation}
\label{eq:reghs}
R_{hs}(g,p) = \sum_{\mathbf{y}} \left\|(\mathbf{H}\ast g)(\mathbf{y})\right\|_{S(p)},
\end{equation}
where $\left\|(\cdot)\right\|_{S(p)}$ is Schatten norm of order $p=[1,\infty]$, which is essentially $l_p$ norm of the Eigen values of its matrix argument.  The above cost can be re-written as
\begin{equation}
\label{eq:reghs2}
R_{hs}(g,p) = \sum_{\mathbf{y}}
	\left[
	|\mathcal{E}_1((\mathbf{H}\ast g)(\mathbf{y}))|^p + 
	|\mathcal{E}_2((\mathbf{H}\ast g)(\mathbf{y}))|^p \right]^{1/p}
\end{equation}
where $\medmath{\mathcal{E}_1(\cdot)}$  and $\medmath{\mathcal{E}_2(\cdot)}$ denote the operators that return the Eigen values of the matrix argument.  From the expression for directional derivative given above,  and from the fact that the Eigen vectors of a symmetric matrix are orthogonal,  it can be shown that 
\begin{align}
\mathcal{E}_1((\mathbf{H}\ast g)(\mathbf{y})) & = 
\mathlarger{{\partial}}_{ \bar{\mathbf{g}}_1(\mathbf{y}) 
		\bar{\mathbf{g}}_1(\mathbf{y}) }g(\mathbf{y}) = \\ \nonumber 
& (\bar{\mathbf{g}}_1(\mathbf{y}))^T \left[(\mathbf{H}\ast g)(\mathbf{y})\right]
	\bar{\mathbf{g}}_1(\mathbf{y}), \;\; \mbox{and} \\
\mathcal{E}_2((\mathbf{H}\ast g)(\mathbf{y})) & = 
\mathlarger{{\partial}}_{ \bar{\mathbf{g}}_2(\mathbf{y}) 
		\bar{\mathbf{g}}_2(\mathbf{y}) }g(\mathbf{y}) = \\ \nonumber
 &(\bar{\mathbf{g}}_2(\mathbf{y}))^T \left[(\mathbf{H}\ast g)(\mathbf{y})\right]
	\bar{\mathbf{g}}_2(\mathbf{y}),
\end{align}
where $\bar{\mathbf{g}}_1(\mathbf{y})$  and $ \bar{\mathbf{g}}_2(\mathbf{y})$ are the Eigen vectors of $(\mathbf{H}\ast g)(\mathbf{y})$.  In other words, the Eigen values of $(\mathbf{H}\ast g)(\mathbf{y})$   are the directional derivatives of $g(\mathbf{y})$  taken along the Eigen directions of the Hessian.  It can also be shown that the cross derivative satisfy $\mathlarger{{ \partial} }_{ \bar{\mathbf{g}}_1(\mathbf{y})\bar{\mathbf{g}}_2(\mathbf{y}) }u(\mathbf{y}) = (\bar{\mathbf{g}}_1(\mathbf{y}))^T \left[(\mathbf{H}\ast g)(\mathbf{y})\right] \bar{\mathbf{g}}_2(\mathbf{y}) = 0$.

\subsection{Maximum entropic probability density on directional derivatives}

The main goal here to is develop an improved form  for the prior probability, $p(g,v, k)$, where $g$ is the  underlying image that generated  the measurement,  and $v$ is prior estimate of the required image, which we call the structure guide.  The idea is to build this prior probability on the distribution of Eigen values of the  Hessian of $g$  in a spatially adaptive manner,  where the information for spatial adaptation is extracted from $v$.  To this end, we define the directional derivatives applied on image $v$ with directions specified by the Eigen vectors of the Hessian of $g$,   as given below:
\begin{align}
\label{eq:d1g2vdef}
\mathcal{D}_{1,g}(v(\mathbf{y})) & = \mathlarger{{\partial}}_{ \bar{\mathbf{g}}_1(\mathbf{y}) 
		\bar{\mathbf{g}}_1(\mathbf{y}) }v(\mathbf{y}) \\ \nonumber &= 
(\bar{\mathbf{g}}_1(\mathbf{y}))^T \left[(\mathbf{H}\ast v)(\mathbf{y})\right]
\bar{\mathbf{g}}_1(\mathbf{y}), \\
\label{eq:d2g2vdef}
\mathcal{D}_{2,g}(v(\mathbf{y})) & = \mathlarger{{\partial}}_{ \bar{\mathbf{g}}_2(\mathbf{y}) 
		\bar{\mathbf{g}}_2(\mathbf{y}) }v(\mathbf{y})\\ \nonumber & = 
\medmath{(\bar{\mathbf{g}}_2(\mathbf{y}))^T \left[(\mathbf{H}\ast v)(\mathbf{y})\right]
	\bar{\mathbf{g}}_2(\mathbf{y})}, 
\end{align}
With the above notation, we can observe that 
\begin{align}
\label{eq:d1e1}
\mathcal{D}_{1,g}(g(\mathbf{y})) & =  
\mathcal{E}_{1}((\mathbf{H}\ast g)(\mathbf{y})), \\
\label{eq:d2e2}
\mathcal{D}_{2,g}(g(\mathbf{y})) & = 
\mathcal{E}_{2}((\mathbf{H}\ast g)(\mathbf{y})).
\end{align}
 Now our goal can be stated as to build prior probability model for the distribution of $\mathcal{D}_{1,g}((g(\mathbf{y}))$  and $\mathcal{D}_{2,g}((g(\mathbf{y}))$ based on the prior information  present in $\mathcal{D}_{1,g}((v(\mathbf{y}))$  and $\mathcal{D}_{2,g}((v(\mathbf{y}))$. To this end, we make the following  hypotheses:
\begin{itemize}
	\item
	{\bf [H1]}: For each pixel location $\mathbf{y}$, $\mathcal{D}_{i,g}(v(\mathbf{y}))$ will be a estimate of mean for $\mathcal{D}_{i,g}(g(\mathbf{y}))$. This is justified, because,  the derivative operators are linear operators,  $v$ is smoothed version of the required solution $g$.
	\item
	{\bf [H2]}: A larger value of $|\mathcal{D}_{i,g}(v(\mathbf{y}))|$ indicates a smaller spatial support of features contributing to it.  This means that---if  $\mathcal{D}_{i,g}(v(\mathbf{y}))$ is approximately related to $\mathcal{D}_{i,g}(g(\mathbf{y}))$ by smoothing with a window---there is a larger uncertainty on the exact location of such features within this windows.   Hence,  provided that $\mathcal{D}_{i,g}(v(\mathbf{y}))$ is an unbiased  estimate of mean for $\mathcal{D}_{i,g}(g(\mathbf{y}))$,   
	$|\mathcal{D}_{i,g}(v(\mathbf{y}))|$ will also be proportional to the variance of $\mathcal{D}_{i,g}(g(\mathbf{y}))$.  Allowing an user parameter, $q > 0$, we designate that this variance is $k|\mathcal{D}_{i,g}(v(\mathbf{y}))|^{q}$, where $k$ is a proportionality constant.  
	\item
	{\bf [H3]}: The closeness of $v$ to the required solution $g$ determines the magnitude of $k$.  Higher closeness will corresponds lower magnitude of $k$. 
\end{itemize}

It should be emphasized that the second hypothesis does  not mean that the local variance of 
 $\mathcal{D}_{i,g}(g(\mathbf{y}))$  within a neighborhood around $\mathbf{y}$ is given by
  $\mathcal{D}_{i,g}(v(\mathbf{y}))$.  It actually means the following:    for any given value $z$,
  if $\{\mathbf{y}_j\}_{j=1}^M$ is the list of points such that  
  $\mathcal{D}_{i,g}(v(\mathbf{y}_j))=z, j=1,\ldots,M$,  then the variance of the sample set
   $\{\mathcal{D}_{i,g}(g(\mathbf{y}_j)), j=1,\ldots,M\}$  is proportional to 
   $|z|^q$.   Proving the validity of these hypotheses is beyond the
 scope this paper. However,  the reconstruction result obtained by means of the regularization 
 developed using these hypotheses will demonstrate the validity implicitly.
 
As we know the mean and variance of distribution of $\mathcal{D}_{i,g}(g(\mathbf{y}))$, we use the maximum entropy principle for determine the prior probability,  which says that, given the information derived from the data,  the best prior probability distribution that will lead to minimal distortion is the one that has the highest entropy~\cite{jaynes1957information1, jaynes1957information2}.  Further it is also known that,  if the variance and mean are known,  the probability model that has the highest entropy is Gaussian model.  Hence,  the proposed probability density function is given by
\begin{equation}
p(g,v,k) = a \prod_\mathbf{y} \exp\left(- \frac{1}{k} \sum_{i=1}^2
	\frac{(\mathcal{D}_{i,g}(g(\mathbf{y}))-
		\mathcal{D}_{i,g}(v(\mathbf{y})))^2}
	{|\mathcal{D}_{i,g}(v(\mathbf{y}))|^{q}}\right),
\end{equation}
where $a$ is some normalization constant. Now the main problem in the above form of regularization functional is that, it is hard to minimize, as all the quantities involved are directional derivatives, and directions are dependent on the unknown original image $g$.   Hence,  we propose the following modifications. We first, replace $\mathcal{D}_{i,g}(\cdot)$  by $\mathcal{D}_{i,v}(\cdot)$.  This means essentially that, the directions for the directional derivatives are now obtained from the prior estimate $v$. Hence we write the  prior probability as given below:
\begin{equation}
\label{eq:mer2}
p(g,v,k) = a \prod_\mathbf{y} \exp\left(- \frac{1}{k} \sum_{i=1}^2
	\frac{(\mathcal{D}_{i,v}(g(\mathbf{y}))-
		\mathcal{D}_{i,v}(v(\mathbf{y})))^2}
	{|\mathcal{D}_{i,v}(v(\mathbf{y}))|^{q}}\right),
\end{equation}
Here, $\mathcal{D}_{1,v}((g(\mathbf{y}))$ and $\mathcal{D}_{2,v}((g(\mathbf{y}))$ are similarly defined as defined in the equations \eqref{eq:d1g2vdef} and \eqref{eq:d2g2vdef} with the role of $g$ and $v$ interchanged.  In other words,  these operators are given by
\begin{align}
\label{eq:d1v2udef}
\mathcal{D}_{1,v}(g(\mathbf{y})) & = \mathlarger{{\partial}}_{ \bar{\mathbf{v}}_1(\mathbf{y})\bar{\mathbf{v}}_1(\mathbf{y}) }g(\mathbf{y})\\ \nonumber & = 
(\bar{\mathbf{v}}_1(\mathbf{y}))^T \left[(\mathbf{H}\ast g)(\mathbf{y})\right]\bar{\mathbf{v}}_1(\mathbf{y}), \\
\label{eq:d2v2udef}
\mathcal{D}_{2,v}(g(\mathbf{y})) & = \mathlarger{{\partial}}_{ \bar{\mathbf{v}}_2(\mathbf{y})\bar{\mathbf{v}}_2(\mathbf{y}) }g(\mathbf{y}) \\ \nonumber &= 
(\bar{\mathbf{v}}_2(\mathbf{y}))^T \left[(\mathbf{H}\ast g)(\mathbf{y})\right],\bar{\mathbf{v}}_2(\mathbf{y}).
\end{align}
where $\bar{\mathbf{v}}_1(\mathbf{y})$ and  $\bar{\mathbf{v}}_2(\mathbf{y})$ are
the Eigen vectors $\medmath{(\mathbf{H}\ast v)(\mathbf{y})}$. As directions in the new operators are independent of the unknown image $g$, but obtained from the prior estimate $v$,  the modified regularization functional given in the equation~\eqref{eq:mer2} is easy to minimize. However, the modified cost  poses an another problem.   We observed that this mismatch in the directions lead to artifacts.  This is possibly due to the fact that  the correlation between $\mathcal{D}_{1,v}(( g(\mathbf{y}))$ and $\mathcal{D}_{2,v}(( g(\mathbf{y}))$ is higher than the correlation between  $ \mathcal{D}_{1,g}(( g(\mathbf{y}))$ and $\mathcal{D}_{2,g}(( g(\mathbf{y}))$ due to the mismatch in the directions. To compensate for this, we intend to include the following operator:
\begin{equation}
\label{eq:d12u2vdef}
\mathcal{D}_{1,2,v}(g(\mathbf{y}))  = \mathlarger{{\partial}}_{ \bar{\mathbf{v}}_1(\mathbf{y}) 
		\bar{\mathbf{v}}_2(\mathbf{y}) }g(\mathbf{y}) = 
	(\bar{\mathbf{v}}_1(\mathbf{y}))^T \left[(\mathbf{H}\ast g)(\mathbf{y})\right]
	\bar{\mathbf{v}}_2(\mathbf{y}).
\end{equation}
To incorporate $\mathcal{D}_{1,2,v}(g(\mathbf{y}))$ in the prior probability,  we again need the estimated mean and variance.  As done for $\mathcal{D}_ {1,v} (( g(\mathbf{y}))$ and $\mathcal{D}_{2,v}(( g(\mathbf{y}))$,  we can consider $\mathcal{D}_{1,2,v}(v(\mathbf{y}))$ to be the mean of $\mathcal{D}_{1,2,v}(g(\mathbf{y}))$.  However,  the quantity $\mathcal{D}_{1,2,v}(v(\mathbf{y}))$  is zero and hence it cannot serve for an estimate of variance of $\mathcal{D}_{1,2,v}(( g(\mathbf{y}))$. We propose use  $|\mathcal{D}_{1,v} ((v(\mathbf{y}))| ^{q/2} |\mathcal{D}_{2,v}((v(\mathbf{y}))|^{q/2}$ as the variance for $ \mathcal{D}_{1,2,v}(( g(\mathbf{y}))$. Hence the final form of prior probability can be written as
\begin{align}
\label{eq:maxentp}
p_{me}(g,v,k) &= a \prod_\mathbf{y} \exp\left(-\frac{1}{k} \left(\sum_{i=1}^2
	\frac{(\mathcal{D}_{i,v}(g(\mathbf{y}))-
		\mathcal{D}_{i,v}(v(\mathbf{y})))^2}
	{|\mathcal{D}_{i,v}(v(\mathbf{y}))|^{q}}\right.\right. \\ \nonumber + 
&\left.\left. \frac{(\mathcal{D}_{1,2,v}((g(\mathbf{y})))^2}
	{|\mathcal{D}_{1,v}((v(\mathbf{y}))|^{\frac{q}{2}}|\mathcal{D}_{2,v}((v(\mathbf{y}))|^{\frac{q}{2}}}
	\right)\right),
\end{align}
In the above form, recall that $\medmath{\mathcal{D}_{i,v}(g(\mathbf{y}))}$ is the directional second derivative applied on $g$ along the direction given by the $i$th Eigen vector of Hessian of $v$ at $\mathbf{y}$. On the other hand, $\mathcal{D}_{i,v}(v(\mathbf{y}))$ is the same directional operator applied on $v$ itself,  which is actually the $i$th Eigen value of the Hessian of  $v$ at $\mathbf{y}$.

\subsection{The full cost for reconstruction}
\label{sec:thefullcost}

By following the same convention as that of the MSDA  method,  the negative log of
$p_{me}(\cdot,v,k)$ applied on the candidate image $u$ (minimization variable) becomes the proposed regularization,  referred as the maximum entropic regularization.  In other words, we write,  $-\log p_{me}(u,v,k) =\frac{1}{k}R_{me}(u,v)$. Now the cost for the proposed reconstruction method is obtained by replacing $R_{msda}(u,v)$ by  $R_{me}(u,v)$ in the equation  $\eqref{Eq:MAPcost4}$, which is given below:
\begin{equation}
\label{eq:costmaxent1}
J_{me}(u,v,h, c) = 
	\frac{1}{\sigma^2}
	\sum_{\mathbf{y}}c(\mathbf{y})\left(h(\mathbf{y}) - u(\mathbf{y})\right)^2 
	+ \frac{1}{k}  R_{me}(u,v),
\end{equation}
where 
\begin{align}
\label{eq:regmaxent2}
R_{me}(u,v) &=   
\sum_\mathbf{y} \left[
\sum_{i=1}^2
	\frac{(\mathcal{D}_{i,v}((u(\mathbf{y}))-
		\mathcal{D}_{i,v}((v(\mathbf{y})))^2}
	{|\mathcal{D}_{i,v}((v(\mathbf{y}))|^{q}} +\right. \\ \nonumber
& \left.\frac{(\mathcal{D}_{1,2,v}((u(\mathbf{y})))^2}
	{|\mathcal{D}_{1,v}((v(\mathbf{y}))|^{q/2}|\mathcal{D}_{2,v}((v(\mathbf{y}))|^{q/2}}
\right]
\end{align}
For notational convenience, we rewrite the above cost as given below:
\begin{equation}
\label{eq:costmaxent1n}
J_{me}(u,v,h, c) = 
\sum_{\mathbf{y}}c(\mathbf{y})\left(h(\mathbf{y}) - u(\mathbf{y})\right)^2 
+ \lambda  R_{me}(u,v),
\end{equation}
where $\lambda = \frac{\sigma^2}{k}$.As done is most image reconstruction methods, we will also impose a bound constraint on the solution by modifying the cost as given below,
\begin{equation}
\label{eq:costmaxent2}
\bar{J}_{me}(u,v,h,v) = 
{J}_{me}(u,v,h,c) + \sum_{\mathbf{y}}B(u(\mathbf{y}))
\end{equation}
where  $B(u(\mathbf{y}))$ is an indicator function for the range of allowable pixel values. $B(u(\mathbf{y}))$ has a value of zero if $0 \le u(\mathbf{y}) \le m$ and $\infty$ otherwise, where $m$ is an user-specified upper bound. The above cost can be minimized by a recently developed variant of ADMM method~\cite{ecksteinyao}.

Now we recall that in MSDA, we had the question of optimal    amount of 
smoothing  that relates the  structure guide and required image.   
The new method described above is not based on this  notion of compensation and hence this question is eliminated.   Specifically, in the new method proposed here,  if the prior estimate $v$ is closer to the original image $g$, the proportionality constant, $k$ is smaller as stated in the hypothesis {\bf [H3]}.  Hence, the closeness of the minimum of $\bar{J}_{me}(u,v,h,v)$,  denoted by $u^*$, to the original image, $g$,  is determined by the closeness of prior estimate to $g$. Also,   better the closeness of $v$ to original image,  better will be reconstruction. 

\section{Fractional multi-resolution based reconstruction}

\subsection{Multiresolution reconstruction}

The low resolution estimate, $v$ is, of course,  unknown as well.  However,  we can  get around this problem by using a multiresolution approach as done in the development of MSDA. Suppose that we have a predefined decreasing sequence of integers $\{N_i\}_{i=0}^s$ representing image sizes of a multiresolution pyramid where $N_0\times N_0$ is the size of $c(\mathbf{y})$  and $h(\mathbf{y})$.   Let $\mathcal{L}_{i,j}$ denote, with $j > i$, an upsampling operator that interpolate an image of size $N_j\times N_j$ into an image of size $N_i\times N_i$. We defer the description of the implementation of $\mathcal{L}_{i,j}$ to later.  In $\mathcal{L}_{i,j}$, we will have two possible values for $i$:  either $0$ or $j+1$. $\mathcal{L}_{0,j}$  will be used to define cost functions for level $j$.  $\mathcal{L}_{j+1,j}$ will be used to generate initializations in the multi-resolution scheme as will be explained later. With this,  we define the following cost function with upsampling:
\begin{align}
\label{eq:fullcostme}
\bar{J}_{me}^{(j)}(u,v, c, h)  &= 
\sum_{\mathbf{y}}c(\mathbf{y})\left(h(\mathbf{y}) - (\mathcal{L}_{0,j}u)(\mathbf{y})\right)^2 +  \\ \nonumber &
 \lambda R_{me}(\mathcal{L}_{0,j}u,v) +   \sum_{\mathbf{y}}B((\mathcal{L}_{0,j}u)(\mathbf{y})) 
\end{align}
where  $(\mathcal{L}_{0,j}u)(\mathbf{y})$ denotes accessing the pixel at $\mathbf{y}$  after upsampling of $u$  from  size $N_j\times N_j$ to size $N_0\times N_0$.  Further,  $R_{me}(\mathcal{L}_{0,j}u,v)$,  denotes applying $R_{me}(u,v)$ after upsampling $u$ from  size $N_j\times N_j$ to size $N_0\times N_0$.  We will use the notational convention that $\bar{J}_{me}^{(0)}(u,v, c, h, \lambda) = \bar{J}_{me}(u,v, c, h, \lambda)$. Note that, in $\bar{J}_{me}^{(j)}(u,v, c, h)$,   size of the variable $u$ is $N_j\times N_j$. To define the multi-resolution method, we will also need the following non-adaptive cost function for initializing reconstruction:
\begin{align}
\label{eq:mrqcost}
{J}^{(s)}(u, c, h)  =
\sum_{\mathbf{y}}c(\mathbf{y})\left(h(\mathbf{y}) - (\mathcal{L}_{0,s}u)(\mathbf{y})\right)^2 
+ \lambda \sum_{\mathbf{y}}\left(\sum_{i=1}^3 \left(\left(d_i\ast (\mathcal{L}_{0,s}u)\right)
	(\mathbf{y})\right)^2\right)
\end{align}
Note that ${J}^{(s)}(u, c, h)$  in the above equation is essentially  the quadratic cost given in the equation  \eqref{eq:regrecsda}  except the difference that it is defined through the upsampling $\mathcal{L}_{0,s}$.  In the above equation, $\left(d_i\ast (\mathcal{L}_{0,s}u)\right)(\mathbf{y})$ denotes accessing pixel at location $\mathbf{y}$   after applying convolution by $d_i$ on the upsampled image $\mathcal{L}_{0,s}u$.

With these definitions, we are ready to describe multiresolution approach. 
In the multi-resolution approach, we obtain the initializing reconstruction by
the following minimization:
\begin{equation}
\label{eq:mrr1}
u^{*(s)} = \underset{u}{\operatorname{argmin}} \;\; J^{(s)}(u,  c, h)
\end{equation}
Using $u^{*(s)}$ as initialization,  multi-resolution based reconstruction
involves a series of minimizations for $j= s-1,s-2,\ldots, 0$
as given below:
\begin{eqnarray}
\label{eq:mrr2}
v  = & {\cal L}_{0,j+1}u^{*(j+1)} \\
\label{eq:mrr3}
u^{*(j)}  = & \underset{u}{\operatorname{argmin}} \;\; 
	\bar{J}^{(j)}_{me}(u, v,  c, h)
\end{eqnarray}
Note that for $j=0$, the final minimization,   $u^{*(0)}  =  \underset{u}{\operatorname{argmin}} \;\;\bar{J}^{(0)}_{me}(u,v,h,c)$ gives the required  final reconstruction.

From the reconstruction represented by  the equation  (\ref{eq:mrr1}), we observe the following:  The initial reconstruction $u^{*(s)}$ is obtained as an ${N}_s\times {N}_s$ array,  which is of the smallest size in the pyramid. Here the regularization is not spatially adaptive,  but,  it is a standard quadratic regularization.  However,  since we are computing solution at coarsest resolution,  this is acceptable.   This result is only for  initializing the multiresolution loop. Next,  from the multi-resolution loop represented by the equations (\ref{eq:mrr2})  and (\ref{eq:mrr3}), we observe that, at each step $j$, the reconstruction, $u^{*(j)}$ is obtained by minimizing the cost $\bar{J}^{(j)}_{me}(u, v,  c, h)$ with respect to $u$.  The evaluation of the cost, $\bar{J}^{(j)}_{me}(u, v,  c, h)$, on the minimization variable is actually carried out via  upsampling by ${\cal L}_{0,j}$ from size $N_{j}\times N_{j}$ to size $N_0\times N_0$.  The structure guide for this cost is obtained from previous reconstruction, $u^{*(j+1)}$, via upsampling by  ${\cal L}_{0,j+1}$ from size $N_{j+1}\times N_{j+1}$ to size $N_0\times N_0$.  As the loop progresses,  the structure guide improves,  which improves the maximum entropic regularization, which in turn becomes the structure for the next reconstruction and so on. Clearly, at the end of the loop, the reconstruction, $\medmath{u^{*(0)}}$ will have less artifacts than the reconstruction obtained by standard $\ell_1$ regularization,  because of this multiresolution based adaptive regularization. It should be noted that, at each step $j$,  the minimization variable as well as the result of minimization, $u^{*(j)}$,  is of size $N_j\times N_j$. Note that the step specified by the equation \eqref{eq:mrr3} itself requires iterative  computation, which needs an initialization.  An efficient initialization can be obtained from $u^{*(j+1)}$ by upsampling operator ${\cal L}_{j,j+1}$, which generates an  $N_{j}\times N_{j}$ image from an $ N_{j+1}\times N_{j+1}$.

It should be emphasized, in the definition of $\bar{J}^{(j)}_{me}(u,v, c,h)$, we have used the upsampler ${\cal L}_{0,j}$---which interpolates from size $N_{j}\times N_{j}$ to size $N_0\times N_0$---for both the data fitting and regularization parts.  Note that the interpolation is essential only for the data fitting part, and the regularizer can be defined directly on the $N_{j}\times N_{j}$ image variable, $u$.  However, from our trials, we found that the current implementation  produces better results.  In a particular,  the upsampled minima $\{{\cal L}_{0,j} u^{*(j)}\}_{j=0}^s$ turn out to be better approximations to the original image $g$ that generated the measured image $h$.  This is important because, at each step $j$, ${\cal L}_{0,j+1} u^{*(j+1)}$ works as the structure guide for obtaining the reconstruction $u^{*(j)}$.

\subsection{Fractional multiresolution}

Note that the image size in most of the multiresolution schemes used so far in the literature satisfy  $\frac{N_j}{N_{j+1}}=2$. We propose that,  for better reconstruction results, this ratio should be a rational number in the open interval $(1,2)$.  We will first justify the need for this, and then explain the specific form of the sequence $\{N_j\}_{j=0}^s$ and the implementation of the upsampling operators $\{\mathcal{L}_{i,j}: 0 \le i,j \le s; i  < j\}$.  

We first note that,  in the sequence of minimization results in the multi-resolution, $\{u^{*(j)}\}_{j=0}^s$,   the reconstruction corresponding to  $j=0$, $u^{*(0)}$ is the final required reconstruction.   The other reconstructions for $j > 0$ can be considered to be the approximations to the final reconstruction; we consider that the actual approximations are given by the upsampled versions of these reconstructions, namely $\{ \mathcal{L}_ {0,j} u^{*(j)}\}_{j=0}^s$.  Now, let $g^{*(j)}$ denote  the best approximation for $g$, in sense of some matching criterion between $\medmath{ \mathcal{L}_{0,j}g^{*(j)}}$ and   $g$.  Now,  by applying argument given at the end of Section \ref{sec:thefullcost}, we can say that  the closeness  of $\mathcal{L}_{0,j}u^{*(j)}$ to $\mathcal{L}_{0,j}g^{*(j)}$  is determined by the closeness of the structure guide used in the reconstruction which is  $\mathcal{L}_{0,j+1}u^{*(j+1)}$. This in turn is determined by how close is $\mathcal{L}_{0,j+1}g^{*(j+1)}$ to $\mathcal{L}_{0,j}g^{*(j)}$. This suggests that the ratio $\frac{N_j}{N_{j+1}}$ should be sufficiently low.   However, it should not be too low of course; otherwise $s$ has be a large number meaning that we will need too many steps in the multiresolution loop. 

For the convenience of implementation,  we consider the size to be of the form $N_j = n_sN_d + (s-j)N_d$, for some positive integers $n_s$ and $N_d$ such that, for any $i$ and $j$ in $[0,s]$, the ratio $\frac{N_i}{N_j}$  will be a rational number. Hence,  the operator $\mathcal{L}_{i,j}$ with $j > i$, which denotes upsampling from size $N_j\times N_j$ to size $N_i\times N_i$,  is rational upsampling operator. If $\frac{N_j}{N_i} = \frac{L_j}{D_i}$ where $L_j$ and $D_i$ are some integer with no common factors,  then $\medmath{ \mathcal{L}_{i,j}}$  can be implemented by an $L_j$-fold upsampling  followed by a  $D_i$-fold  downsampling.  This can be implemented using the following three steps \cite{bssp1,bssp2}:(i) expansion by a factor of $L_j$ along both axes which is essentially inserting  $L_j-1$ zeros for every sample along both axes; (ii) convolve with filter $\frac{1}{L_j}(1+z^{-1})^{L_j}$ along both axes; (iii) decimation by a factor of $D_i$  which is discarding $D_i-1$ samples for every block of $D_i$ samples.

\section{Experimental results}
\label{sect:experimental}
Confocal microscope is a $3$D imaging modality in which a series of point wise scanned $2$D images are stacked together.  However,  it is often applied in $2D+t$ mode to observe fast cellular processes.  In this context, our goal is to suggest the proposed method,  which we name Maximum Entropic Regularized Reconstruction (MERR), as computationaltool for optimally trading resolution against acquisition time, by acquiring point measurements only at randomly sub-sampled incomplete set of locations.  We also claim that our computational tool allows to reduce photo-toxicity because of its robustness against noise.  To demonstrate both claims, we generate two type test data as given below.
\begin{itemize}
	\item 
	We measure regular grid 2D images from  a fixed sample, where one is acquired with full excitation intensity,  and others are acquired with different lower excitation intensities.  Different sample sets obtained by randomly selecting points from the low excitation images are used as test inputs for MERR, and the full 2D image corresponding to highest exposure intensity is used as the ground truth for evaluating the reconstruction quality.
	\item
	We select  few 2D confocal images from Nikon Small World repository as models.  From these models,  we simulate low excitation images by adding mixed Poisson-Gaussian noise, and then generate randomly selected sample sets as inputs for reconstruction. The noiseless models are used as ground truth for evaluating the reconstruction quality.
\end{itemize}

We compare MERR with $l_1$ method~\cite{scherzer1998denoising},  and MSDA method 
\cite{francis2018multiresolution} in terms of structural  similarity (SSIM) \cite{wang2004image} of the reconstructed image with respect to the ground truth. Note that MERR is different from MSDA in two ways: (i)  In MERR, the regularization is constructed using multiresolution based directionally adaptive filters by means of a maximum entropy formulation;  on the other hands, in MSDA,  the regularization is in an isotropic form  with a multiresolution based reweighting; (ii) in MERR,  we use fractional multiresolution,  whereas in MSDA, standard multiresolution is used.  
We do not compare with the $\ell_2$ regularization here because, we have already 
demonstrated the superiority of MSDA  over $\ell_2$ 
\cite{francis2018multiresolution}.

Since, the main goal here is to demonstrate the efficiency of maximum entropic regularization, we use the ground truth itself to tune the smoothing parameters for all the methods, and  $q$ for the proposed method, MERR.  While using the ground truth for tuning the smoothing parameter is common in the literature~\cite{lefkimmiatis2013hessian,lefkimmiatis2012hessian},  we need extra tuning effort for determining $q$. However,  in our experiments, the optimal value for $q$ was found independent of noise level and sample density,  and was dependent only on the structure of the image. Hence, for  a practitioner applying our method, the values of $q$ can be kept fixed as long as the experiment  involves the same organelle.  This situation is similar to that of MSDA, whose parameters $q$ and $r$ were dependent only on the nature of image structure \cite{francis2018multiresolution}. For determining optimal value for $q$, we performed grid search in the range $0.5-1.0$  with
step size  $0.1$. For  implementing $\medmath{|\mathcal{D}_{i,v} ((v(\mathbf{y}))|^q}$ and $\medmath{|\mathcal{D}_{i,v}((v(\mathbf{y}))|^{q/2}}$, we used the approximations $\medmath{(\epsilon+|\mathcal{D}_{i,v} ((v(\mathbf{y} ))|)^q}$ and $\medmath{(\epsilon+|\mathcal{D}_{i,v}((v(\mathbf{y}))|)^{q/2}}$  with $\epsilon = 10^{-6}$ to ensure differentiability. For all reconstructions performed by MERR,  we set the parameters such that $\medmath{N_d=16}$ and $\medmath{N_0/N_s = 4}$.   While making $\medmath{N_d}$ smaller will always  improve the reconstruction quality, the improvement was found to be insignificant;  on the other hand, the rise in the computational  complexity becomes  unaffordable.  Similarly, in principle, making $\medmath{N_0/N_s}$ larger can improve the reconstruction quality, but,  it did not significantly improve the reconstruction quality.

\subsection{Experiments on real measured images}     

\begin{figure*}[ht]
				\centering
				\includegraphics[width=0.95\linewidth]{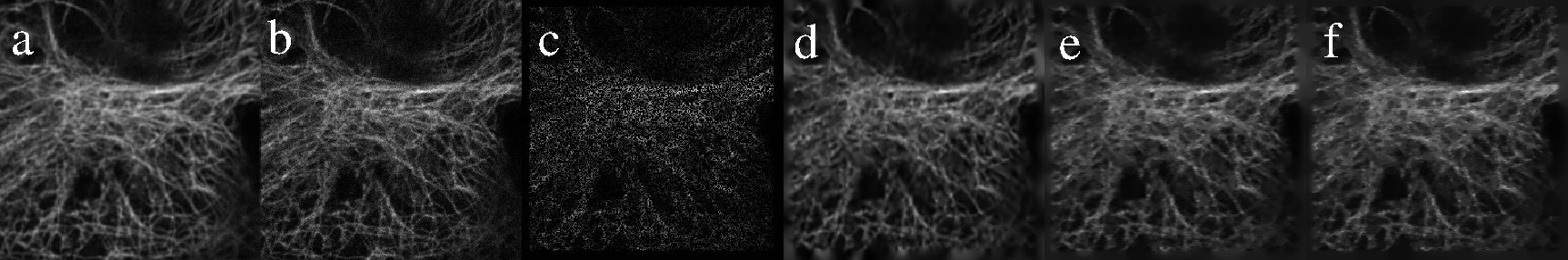}
				\caption{a) Reference Vero Cell image acquired with full laser power;  b) Vero cell image acquired with $40\%$
					laser power; c) Nonuniformly sampled Vero cell image obtained from (b) with $40 \%$ subsampling; d) Reconstruction obtained by  MERR method from (c);  e) Reconstruction obtained by  MSDA from (c);  
					f) Reconstruction obtained by  $\ell_1$  method from (c).}
				\label{Result_combined}
			\end{figure*}

For our first experiment, we imaged Vero cell labeled with Abberior Star $580$ dye with imaging region of size of $52.2\mu m\times 51.2\mu m$  comprising of $512\times 512$ pixels.  The images were taken with  five levels of excitation intensities; the highest excitation intensity was set such that the image is nearly noise-free, which will be used as the ground truth.  Other reduced intensity levels were chosen to be $10 \%, 20 \%, 30 \%$, and  $40 \%$ of the highest.  Figure~\ref{Result_combined}.a shows the $256\times 256$  cropped view of the image acquired at the full laser power level and~\ref{Result_combined}.b shows the cropped $256\times 256$ image acquired at $40\%$ laser power level.  From each of the  images corresponding to these four reduced levels of excitation, we extract five sample set with densities $30 \%, 35 \%, 40  \%, 45 \%$, and  $50 \%$. This makes a  total of $20$ test datasets.  The reconstruction results from these $20$ datasets are compared in  table~\ref{Table_original2} and~\ref{Table_original1}. From the tables, it is clear that the proposed approach, MERR,  outperforms both $\ell_1$ and MSDA  methods. Moreover, it can be observed that, for MERR, the optimal value of $q$  remains constant  independent of noise level and the sample density, and is  equal to $0.9$.      Moreover, the improvement yielded by MERR   increases as the noise level increases.  Among the compared methods, $\ell_1$ method is the   most sensitive one to noise  yielding the lowest SSIM score.
	
{\renewcommand{\arraystretch}{0.8}			
	\begin{table*}[ht]
		\scalebox{0.85} {
\begin{tabular}{|c|c|c|c|c|c|c|c|c|c|}
\hline
\multirow{3}{*}{\begin{tabular}[c]{@{}c@{}}Percentage \\ of laser\\ power used\end{tabular}} & \multicolumn{9}{c|}{Sampling Densities}                                                                                 \\ \cline{2-10} 
                                                                                             & \multicolumn{3}{c|}{30}  & \multicolumn{3}{c|}{35}  & \multicolumn{3}{c|}{40}                                           \\ \cline{2-10} 
                                                                                             & MERR  & MSDA  & $\ell_1$ & MERR  & MSDA  & $\ell_1$ & MERR  & \multicolumn{1}{l|}{MSDA} & \multicolumn{1}{l|}{$\ell_1$} \\ \hline
40                                                                                           & 0.792 & 0.771 & 0.737    & 0.811 & 0.788 & 0.760    & 0.826 & 0.800                     & 0.772                         \\ \hline
30                                                                                           & 0.783 & 0.759 & 0.727    & 0.794 & 0.770 & 0.742    & 0.807 & 0.780                     & 0.753                         \\ \hline
20                                                                                           & 0.753 & 0.731 & 0.697    & 0.771 & 0.745 & 0.716    & 0.779 & 0.754                     & 0.726                         \\ \hline
10                                                                                           & 0.714 & 0.682 & 0.657    & 0.730 & 0.706 & 0.679    & 0.742 & 0.717                     & 0.689                         \\ \hline
\end{tabular}}
	\caption{Comparison of  SSIM score for  $\ell_1$, MSDA, and MERR reconstructions for 
				sampling densities from $30\%$  to $40\%$ of $256\times256$ images acquired at various laser power 
				levels on the Vero Cell image (figure~\ref{Result_combined})}
			\label{Table_original2}
\end{table*}}

{\renewcommand{\arraystretch}{0.8}			
	\begin{table*}[ht]
		\scalebox{0.85} {
\begin{tabular}{|c|c|c|c|c|c|c|}
\hline
\multirow{3}{*}{\begin{tabular}[c]{@{}c@{}}Percentage \\ of laser\\ power used\end{tabular}} & \multicolumn{6}{c|}{Sampling Densities}             \\ \cline{2-7} 
                                                                                             & \multicolumn{3}{c|}{45}  & \multicolumn{3}{c|}{50}  \\ \cline{2-7} 
                                                                                             & MERR  & MSDA  & $\ell_1$ & MERR  & MSDA  & $\ell_1$ \\ \hline
40                                                                                           & 0.831 & 0.805 & 0.779    & 0.837 & 0.812 & 0.790    \\ \hline
30                                                                                           & 0.814 & 0.784 & 0.763    & 0.822 & 0.800 & 0.774    \\ \hline
20                                                                                           & 0.788 & 0.762 & 0.735    & 0.800 & 0.774 & 0.749    \\ \hline
10                                                                                           & 0.749 & 0.722 & 0.695    & 0.755 & 0.731 & 0.708    \\ \hline
\end{tabular}}
	\caption{Comparison of  SSIM score for  $\ell_1$, MSDA, and MERR reconstructions for 
				sampling densities from $45\%$ to $50\%$ of $256\times256$ images acquired at various laser power 
				levels on the Vero Cell image (figure~\ref{Result_combined})}
			\label{Table_original1}
\end{table*}}

\begin{figure*}[ht]
	\centering
	\includegraphics[width=0.95\linewidth]{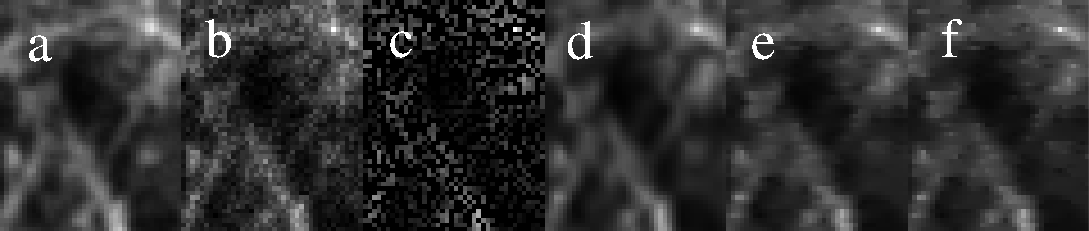}
	\caption{Zoomed in view of figure \ref{Result_combined} (physical size: 
		$5\;\mu m \times 4\;\mu m$) (a) Reference Vero Cell image acquired 
		with full laser power;  b) Vero cell image acquired with $40\%$
		laser power; c) Nonuniformly sampled Vero cell image obtained from (b) with $40 \%$ subsampling; 
		d) Reconstruction obtained by  MERR method from (c);  e) Reconstruction obtained by  MSDA from (c);  
		f) Reconstruction obtained by  $\ell_1$  method from (c).}
	\label{Result_crop}
\end{figure*}

To visually demonstrate the quality of reconstruction, we have provided the reconstruction results from $40\%$ of samples taken from the image acquired at $40\%$ laser power in the figure~\ref{Result_combined}. Figure~\ref{Result_combined}.a shows the image obtained by imaging the Vero cell with full lase intensity, which we  designate as the ground truth image. Figure~\ref{Result_combined}.b represents the image acquired at $40\%$ laser power level.  From this $40\%$ samples were drawn randomly to get the nonuniformly sampled image, which is shown in figure~\ref{Result_combined}.c. Reconstructions   obtained  from this dataset using MERR, MSDA and $\ell_1$ methods are given in the figures \ref{Result_combined}.d,  \ref{Result_combined}.e, and  \ref{Result_combined}.f respectively. From the results, it is clear that the proposed method yield better reconstruction than the existing ones. Moreover, the proposed method  is better  suited to preserve the structures in the image which is responsible for the improvement in quality  of reconstruction measured by the SSIM score. To highlight this fact, we have provided a zoomed in  view of the reconstruction  in the figure \ref{Result_crop}.  From the figure, it is clear that $\ell_1$ reconstruction has largest amount of artifacts in form of spikes.  MSDA has reduced amount of artifacts,  but the artifacts are still significant. On the other hand, there are no visible artifacts in the MERR reconstruction. It should be emphasized that there is a loss of resolution in reconstructed images of all three methods, which is inevitable because of noise and subsampling;  however,  the main factor that makes MERR reconstruction superior is the absence of spiky artifacts.

\subsection{Experiments on images with simulated  noise}

\begin{figure}
	\centering
	\includegraphics[width=0.95\linewidth]{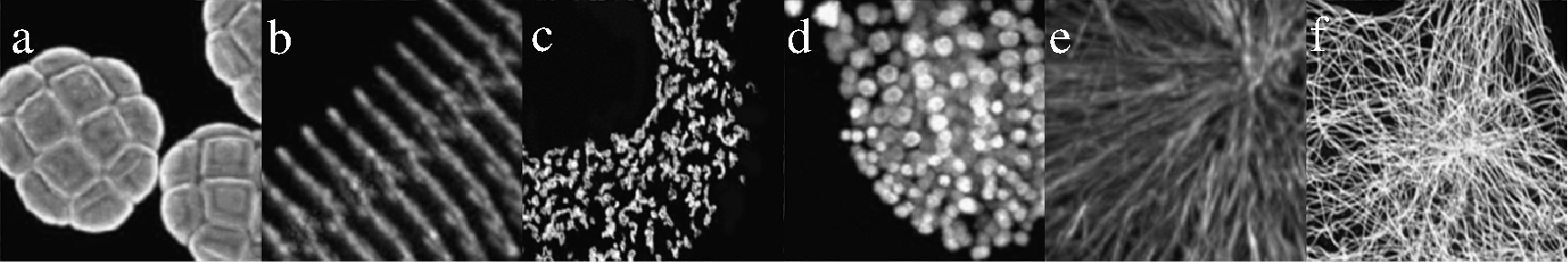}
	\caption{ Test images from Nikon Small World a) Acacia dealbata pollen grains (Img1);  
		b) Human cardiac myocytes (Img2); c) Golgi complex (Img3); d) Thale cress root (Img4); 
		e) Epithelial cell in anaphase (Img5); f) Tubulin image (Img6).}
	\label{Test_images}
\end{figure}

\begin{figure*}[ht]
	\centering 
	\includegraphics[width=0.95\linewidth]{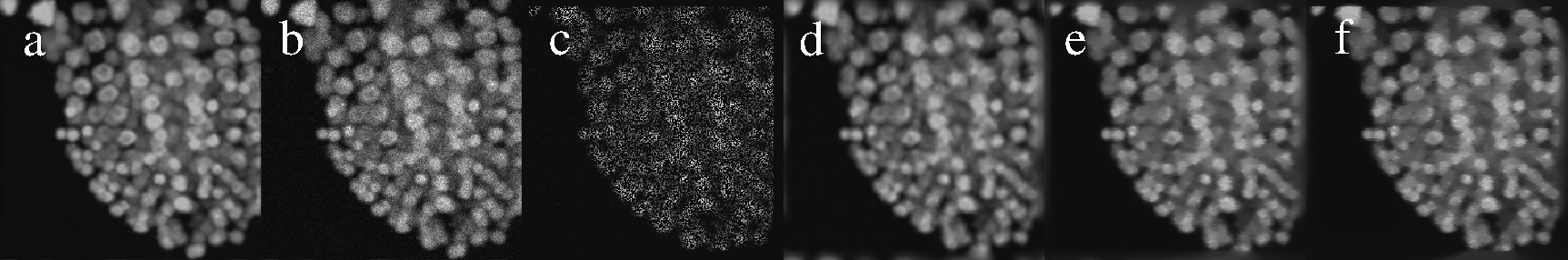}
	\caption{ Comparison of reconstruction results obtained from 
		$40 \%$ of random samples 
		from Thale cress root image (Figure ~\ref{Test_images}.d) with $14.34$dB SNR.
		(a)  ground truth; 
		(b) noisy image with input SNR $14.34$dB; 
		(c)   $40\%$ of samples taken from  (b); 
		(d)  reconstruction of  MERR from (c); 
		(e) reconstruction of MSDA from (c); 
		(f) reconstruction of $\ell_1$ method from (c).}
	\label{Test_imageResult}
\end{figure*}

\begin{figure*}[ht]
	\centering
	\includegraphics[width=0.95\linewidth]{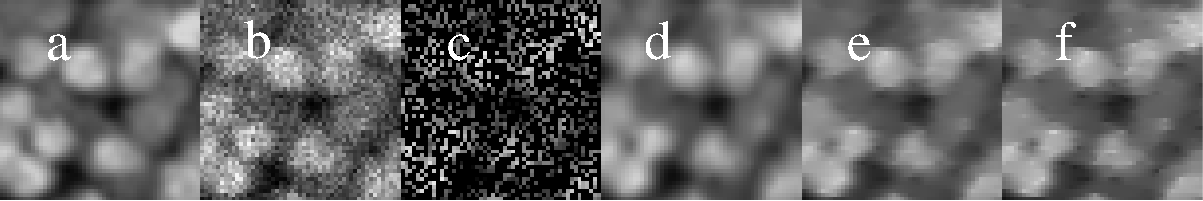}
	\caption{ Zoomed-in view of figure \ref{Test_imageResult} (physical size: 
		$5.1\;\mu m \times 5.1\;\mu m$).
		(a)  ground truth; 
		(b) noisy image with input SNR $14.34$dB; 
		(c)   $40\%$ of samples taken from  (b); 
		(d)  reconstruction of  MERR from (c); 
		(e) reconstruction of MSDA from (c); 
		(f)  reconstruction of $\ell_1$ method from (c).}
	\label{CropTest_imageResult}
\end{figure*}

The goal here is to compare MERR with other methods on images  having a variety of structures. For this purpose, we selected 6 images from Nikon Small World repository, which are displayed in the  figure \ref{Test_images}.  We have chosen the set such that the images have different types of distribution of derivative values. The Tublin image in~\ref{Test_images}.f is similar to the Vero cell image with identical structures and dense nonzero derivative coefficients. However, the Golgi complex in~\ref{Test_images}.c has a sparse distribution of nonzero derivative values. All the remaining images in the figure have the derivative distribution  in between these two images. Hence, the selected set is  a good representation for the biological images in the viewpoint of testing the suitability of a new regularization scheme. 
These images are considered as the ground truth model,  and to simulate measurements with varying excitation intensities, we added varying levels of noise.  We chose three levels of noise such that the visual perception of the noise level matches with noise level of images with $10 \%, 30 \%$ and $40 \%$   
excitation intensities in the previous experiments.  The corresponding SNRs turn out to be $12.10$ dB, $13.34$ dB, and $14.34$ dB respectively. From each of these noisy images, we selected randomly selected sample sets with densities $30 \%, 40 \%$ and $50 \%$. This  makes a total of $6\times 3 \times 3 = 54$ test data sets.  Reconstruction results of various methods applied on these set are compare in the table \ref{Table_Test1}. From the table, it is clear that the relative performance of various methods are in the same order, and MERR outperforms other methods.  In figure \ref{Test_imageResult}, we display images reconstructed from $40 \%$ of random samples from Thale cress root image (Figure ~\ref{Test_images}.d) with $14.34$dB SNR.  Here too, it is clear that the proposed approach clearly outperforms the  competing  methods.  A  zoomed-in view of the comparison is given in the figure \ref{CropTest_imageResult}, where it is evident that the reconstruction from MERR has least amount of artifacts, and $\ell_1$ method has the highest amount of artifacts.

{\renewcommand{\arraystretch}{0.8}
	\begin{table*}[ht]
		\begin{center}
			\scalebox{0.85}
			{
				\begin{tabular}{|c|c|c|c|c|c|c|c|c|c|c|}
					\hline
					\multirow{3}{*}{Images} & \multirow{3}{*}{\begin{tabular}[c]{@{}c@{}}Input \\ SNR (dB)\end{tabular}} & \multicolumn{9}{c|}{Sampling Densities}                                           \\ \cline{3-11} 
					&                                                                            & \multicolumn{3}{c|}{30\%} & \multicolumn{3}{c|}{40\%} & \multicolumn{3}{c|}{50\%} \\ \cline{3-11} 
					&                                                                            & MERR  & MSDA  & $\ell_1$ & MERR  & MSDA  & $\ell_1$ & MERR  & MSDA  & $\ell_1$ \\ \hline
					\multirow{3}{*}{Img1}   & 14.34                                                                      & 0.871  & 0.839 & 0.799    & 0.884  & 0.857 & 0.821    & 0.898  & 0.871 & 0.842    \\ \cline{2-11} 
					& 13.34                                                                      & 0.861  & 0.832 & 0.790    & 0.875 & 0.845 & 0.807    & 0.884  & 0.856 & 0.807    \\ \cline{2-11} 
					& 12.10                                                                      & 0.838  & 0.811 & 0.770    & 0.862  & 0.818 & 0.784    & 0.873  & 0.834 & 0.816    \\ \hline
					\multirow{3}{*}{Img2}   & 14.34                                                                      & 0.934  & 0.903 & 0.884    & 0.937  & 0.907 & 0.888    & 0.941  & 0.913 & 0.894    \\ \cline{2-11} 
					& 13.34                                                                      & 0.926  & 0.895 & 0.871    & 0.933  & 0.902 & 0.879    & 0.933  & 0.903 & 0.880    \\ \cline{2-11} 
					& 12.10                                                                      & 0.912  & 0.869 & 0.841    & 0.916  & 0.879 & 0.851    & 0.920  & 0.895 & 0.869    \\ \hline
					\multirow{3}{*}{Img3}   & 14.34                                                                      & 0.902  & 0.870 & 0.840    & 0.918  & 0.895 & 0.870    & 0.924  & 0.906 & 0.890    \\ \cline{2-11} 
					& 13.34                                                                      & 0.885  & 0.853 & 0.810    & 0.891  & 0.871 & 0.862    & 0.911  & 0.889 & 0.883    \\ \cline{2-11} 
					& 12.10                                                                      & 0.861  & 0.842 & 0.800    & 0.886  & 0.859 & 0.833    & 0.898  & 0.860 & 0.858    \\ \hline
					\multirow{3}{*}{Img4}   & 14.34                                                                      & 0.888  & 0.857 & 0.834    & 0.902  & 0.873 & 0.857    & 0.909  & 0.885 & 0.857    \\ \cline{2-11} 
					& 13.34                                                                      & 0.879  & 0.847 & 0.842    & 0.896  & 0.868 & 0.845    & 0.899  & 0.865 & 0.851    \\ \cline{2-11} 
					& 12.10                                                                      & 0.866  & 0.826 & 0.803    & 0.879  & 0.845 & 0.823    & 0.889  & 0.854 & 0.843    \\ \hline
					\multirow{3}{*}{Img5}   & 14.34                                                                      & 0.853  & 0.831 & 0.763    & 0.871  & 0.822 & 0.807    & 0.885  & 0.837 & 0.817    \\ \cline{2-11} 
					& 13.34                                                                      & 0.841  & 0.811 & 0.750    & 0.855  & 0.845 & 0.794    & 0.884  & 0.856 & 0.807    \\ \cline{2-11} 
					& 12.10                                                                      & 0.838  & 0.804 & 0.740    & 0.842  & 0.818 & 0.784    & 0.873  & 0.834 & 0.792    \\ \hline
					\multirow{3}{*}{Img6}   & 14.34                                                                      & 0.798  & 0.725 & 0.682    & 0.834  & 0.764 & 0.727    & 0.854  & 0.791 & 0.761    \\ \cline{2-11} 
					& 13.34                                                                      & 0.783  & 0.706 & 0.661    & 0.819  & 0.745 & 0.707    & 0.838  & 0.772 & 0.742    \\ \cline{2-11} 
					& 12.10                                                                      & 0.746  & 0.676 & 0.632    & 0.792  & 0.715 & 0.676    & 0.819  & 0.750 & 0.718    \\ \hline
				\end{tabular}
			}
			\caption{Comparison of SSIM scores for  $\ell_1$, MSDA, and  MERR methods from samples generated 
				from the images given in the figure \ref{Test_images}. }
			\label{Table_Test1}
		\end{center}
	\end{table*}
}
			
\section{Conclusions}
\label{sec:conclusion}
 
We addressed the problem of reconstructing 2D images from randomly undersampled noisy confocal microscopy samples.  While quadratic regularization functional that were originally used for this type of problems tend to over-smooth the reconstructed images, total variation regularization functional---which is widely is used in solving inverse problems---results in artifacts in the reconstruction. We developed a new type of regularization functional  as negative logarithm of probability density function representing the distribution of directional derivatives of required image.   The model for the probability density function
is inferred from a lower resolution estimate of required image based on the maximum entropy principle.  The problem of finding the low resolution estimate of the required image is systematically handled using a multiresolution approach involving a series of regularized reconstruction.  We demonstrated that the proposed regularization method, named maximum entropic regularized reconstruction (MERR), yield significantly improved reconstruction compared to competing methods. Note that, in some case, the improvement in SSIM score is as high as 0.07 which is very significant. Although we did not directly prove the hypotheses used for building our maximum entropic probability  model,  the reconstruction results indirectly demonstrate the validity the hypotheses.

\bibliography{library}

\bibliographystyle{abbrv}

\end{document}